\pdfoutput=1
\newif\ifanonymous
\anonymousfalse
\documentclass[runningheads]{llncs}
\usepackage[T1]{fontenc}
\usepackage[utf8]{inputenc}
\usepackage{amssymb}
\DeclareUnicodeCharacter{2192}{\ensuremath{\rightarrow}}
\DeclareUnicodeCharacter{2212}{\ensuremath{-}}
\DeclareUnicodeCharacter{2265}{\ensuremath{\geq}}
\DeclareUnicodeCharacter{2264}{\ensuremath{\leq}}
\DeclareUnicodeCharacter{2248}{\ensuremath{\approx}}
\DeclareUnicodeCharacter{0394}{\ensuremath{\Delta}}
\DeclareUnicodeCharacter{03C4}{\ensuremath{\tau}}
\DeclareUnicodeCharacter{03A3}{\ensuremath{\Sigma}}
\DeclareUnicodeCharacter{03BA}{\ensuremath{\kappa}}
\DeclareUnicodeCharacter{00B7}{\ensuremath{\cdot}}
\DeclareUnicodeCharacter{00B1}{\ensuremath{\pm}}
\DeclareUnicodeCharacter{2080}{\ensuremath{_0}}
\DeclareUnicodeCharacter{2081}{\ensuremath{_1}}
\DeclareUnicodeCharacter{2082}{\ensuremath{_2}}
\DeclareUnicodeCharacter{2083}{\ensuremath{_3}}
\usepackage{graphicx}
\usepackage{booktabs}
\usepackage{amsmath}
\usepackage{multirow}
\usepackage[hidelinks]{hyperref}
\usepackage{xcolor}
\newcommand{\jev}{Jev}
\newcommand{\flan}{FLAN-UL2}

\begin{document}

\title{Beyond Ranking: Are Decision Models Calibrated Relevance Scorers?}

\ifanonymous
\author{Anonymous Author(s)}
\authorrunning{Anonymous}
\institute{Anonymous Institution(s)}
\else
\author{Shengyao Zhuang\inst{1} \and
Xueguang Ma\inst{2} \and
Matthias Petri\inst{1} \and
Jimmy Lin\inst{2}}
\institute{Keenable\\
\email{\{shengyao,matthias\}@keenable.ai} \and
University of Waterloo, Waterloo, Canada\\
\email{\{x93ma,jimmylin\}@uwaterloo.ca}}
\fi

\maketitle

\begin{abstract}
Decision models such as \jev{} are a new kind of relevance scorer. Their typed \texttt{score} primitive takes a question, an ordered set of textual relevance levels and a state, and returns a probability for each level directly, rather than token likelihoods or generated labels. \jev{}'s vendor promises that these probabilities are calibrated. This motivates revisiting \emph{Beyond Yes and No}, which improved zero-shot pointwise LLM reranking using likelihood-derived scores over fine-grained relevance labels but left calibration against human assessors as future work. In this paper, we first reproduce its open-model \flan{} setting on eight BEIR collections, confirming its ranking trends, and then run \jev{} and two open-weight decision models with the same protocol. \jev{} and the 27B pplx-decider rerank ahead of \flan{} and three cross-encoders on average. We then compare \jev{}'s native probabilities with the human assessors' labels and find them overconfident: of the DL21 pairs it scores above 0.9, only 66\% are relevant. Since this is mostly an error of scale, a two-parameter calibration fitted on 2,257 DL19 pairs removes much of it and transfers to DL21. After calibration, \jev{} has a lower DL21 Brier score than \flan{} and the three cross-encoders, mainly because it separates relevant from non-relevant pairs better. Finally, with calibrated probabilities, choosing each query's cut-off by expected F1 needs no parameter of its own, beats a fixed top-10 cut and matches a cut-off threshold tuned on DL19.

\end{abstract}

\section{Introduction}

Large language models (LLMs) are used both to rank search results~\cite{zuccon2025r2llms} and to judge their relevance~\cite{faggioli2023perspectives}. In both uses, the output is usually reduced to a rank or a label. Much less attention goes to the probability attached to that label. This matters because modern neural networks are often overconfident~\cite{guo2017calibration}, and LLM confidence, whether read from token probabilities or stated in words, is frequently miscalibrated~\cite{tian2023calibration,xiong2024uncertainty}.

In September 2026, TypeSafe released \jev{}, the first of its ``System One'' models, also described as decision models~\cite{typesafe2026jev}. Rather than generating text, it takes a typed question (choice, score or yes/no) about a supplied state and returns a probability for each option. The vendor promises calibrated probabilities and charges well below generative LLM API prices for comparable decisions. Within weeks of its release, \jev{} and open-weight decision models were used as rerankers~\cite{elastic2026jev,fernandes2026jevptbr,tatar2026probrel,dinzinger2026system1,wu2026decisionrank,li2026system1rag} and for evidence selection and claim verification in retrieval-augmented generation~\cite{typesafe2026jev,wu2026sag2,terven2026scanindex,dorosario2026jevrag}. These uses act on \jev{}'s relevance probabilities, yet the independent calibration checks so far~\cite{meng2026audit,emretheus2026jevrag} have not compared them with graded human relevance judgments.

The study \emph{Beyond Yes and No} by Zhuang et al.~\cite{zhuang2024beyond} provides the starting point for testing this promise in reranking tasks. It found that graded relevance labels outperform binary labels, likelihood-derived scores outperform generated labels, and gains flatten beyond three or four levels. It left calibration of these probabilities against human assessors as future work. \jev{} can take the same ordered relevance levels as the study's prompts and return their probabilities directly, so it fits the study's protocol. We therefore reproduce the study and extend it beyond ranking quality to calibration and to the decisions that calibrated probabilities support.

Specifically, we first follow the original study's open-model \flan{} setting and reproduce its BM25 reranking experiments on eight BEIR collections~\cite{thakur2021beir}. On the same candidates, we also run \jev{}, three cross-encoders (MiniLM, bge-reranker-v2-m3 and monoT5-3B) and two open-weight decision models (Cloudflare's 9B Clef-flash and Perplexity's 27B pplx-decider-v1.1). To test whether \jev{}'s probabilities are calibrated, we compare them, along with those of the other models, with human relevance judgments on TREC DL19--21~\cite{craswell2020dl19,craswell2021dl20,craswell2022dl21}, where a passage counts as relevant at grade $\geq 2$. We then fit a two-parameter calibration on DL19 and apply it unchanged to DL20 and DL21 to see whether it transfers to a new year and corpus. Finally, we use the calibrated probabilities to decide where to cut each ranked list.

We find that no model we test, including \jev{}, produces native probabilities that are calibrated against TREC assessors. A two-parameter calibration fitted on DL19 corrects much of this error and transfers to DL21. Under the same calibration, \jev{} has a lower DL21 Brier score than \flan{} and all three cross-encoders, a margin that comes mainly from better separation of relevant and non-relevant passages rather than from a more accurate probability scale. This advantage is not specific to \jev{}. The open 27B pplx-decider matches it after calibration (DL21 Brier 0.144 versus 0.147) and, with the three-level prompt (RG-3L), ranks ahead of it on all eight BEIR collections, whereas the 9B Clef-flash falls to the level of \flan{} on the new corpus. The benefit therefore extends beyond one closed service but varies across decision models.

Once calibrated, the probabilities also support decisions that a ranking score alone cannot, such as how many documents to return for each query. Choosing the depth that maximises F1 estimated from the model's own probabilities needs no tuned cut-off, yet matches a threshold tuned on DL19 (F1 0.638 versus 0.630 for \jev{}) and beats a fixed cut at 10 (0.506). With this rule, \jev{} and pplx-decider also cut better than all three cross-encoders (F1 0.638 and 0.658 versus at most 0.566). This matters for retrieval-augmented generation, where every passage passed to the language model costs context tokens and irrelevant passages can distract it~\cite{lewis2020rag,shi2023distracted}.

Our contributions are:
\begin{itemize}
\item \textbf{Reproduction.} We independently reproduce \emph{Beyond Yes and No} with its open \flan{} model on the eight BEIR collections of the original study. 
\item \textbf{Calibration and its transfer.} We compare native and calibrated outputs with proper scoring rules, separate the accuracy of the probability scale from the ability to rank relevant pairs above non-relevant ones, and test calibrators fitted on one TREC year on a new year and corpus and on four BEIR collections.
\item \textbf{Decisions.} We measure what calibrated probabilities buy for ranked-list truncation: a cut-off rule without a tuned parameter against fixed and tuned cut-offs.
\end{itemize}

\section{Background and Research Questions}
\label{sec:background}

Zero-shot LLM rankers differ in what each prompt asks the model to compare. Pointwise methods score individual documents through relevance generation~\cite{zhuang2024beyond} or query likelihood~\cite{zhuang2023qlm}. Pairwise prompting compares two candidates~\cite{qin2024prp}, setwise prompting compares several~\cite{zhuang2024setwise}, and listwise methods produce permutations, as in RankGPT and RankVicuna~\cite{sun2023rankgpt,pradeep2023rankvicuna}. Fine-tuned approaches include the sequence-to-sequence reranker monoT5~\cite{nogueira2020monot5}, RankLLaMA~\cite{ma2024ranklama}, and Rank-R1, which uses reinforcement learning to learn reasoning before ranking~\cite{zhuang2026rankr1}. We stay within the pointwise family, in which each document receives a score of its own.

\emph{Beyond Yes and No}~\cite{zhuang2024beyond} studies how the response space and score derivation affect pointwise reranking. Its RG-$k$L prompts offer $k$ textual relevance levels; the four-level version uses ``Not Relevant'', ``Somewhat Relevant'', ``Highly Relevant'' and ``Perfectly Relevant''. RG-S$(0,k)$ instead requests an integer rating from $0$ to $k$. Let $s_{i,j}$ be the sequence log-likelihood of label $j$ for query--document pair $i$. For $m$ ordered labels with values $y_j=j$, expected relevance (ER) normalises likelihoods over the allowed labels:
\[
p_{i,j}=\frac{\exp(s_{i,j})}{\sum_{r=0}^{m-1}\exp(s_{i,r})},
\qquad
\mathrm{ER}_i=\sum_{j=0}^{m-1}p_{i,j}y_j.
\]
Peak relevance likelihood (PR) uses $\mathrm{PR}_i=s_{i,m-1}$, the log-likelihood of the highest relevance label. It does not select the most likely label. The Generated variant instead parses the model's generated response and ranks by its relevance level.

The study reranks BM25 top-100 candidates on eight BEIR collections~\cite{thakur2021beir} and reports nDCG@10 using FLAN models. It makes three findings. First, fine-grained labels rank better than binary ones: with \flan{}, the eight-collection average rises from 0.459 for RG-2L to 0.485--0.490 with three or four levels or a 0--2 or 0--4 scale. Second, scoring by label likelihood beats the label the model generates (0.499 against 0.399 for RG-3L with FLAN PaLM2 S). Third, beyond about four levels the gains stop and then fall (for FLAN PaLM2 S, 0.502 at $k=4$ and 0.475 at $k=10$ on the 0--$k$ scale).

However, the paper evaluates ranking only. Its limitations section notes that LLM assessors were not studied, that higher ranking performance does not always mean better relevance calibration, and that turning likelihood-derived scores into calibrated relevance estimates is a non-trivial open direction, with uses such as ranked-list truncation~\cite{zhuang2024beyond,bahri2020choppy}.

\jev{} fits this setting closely. Its \texttt{score} question takes an instruction and an ordered list of textual levels, the same shape as an RG-$k$L prompt, with the query and document in the state. Instead of label likelihoods that have to be normalised, it returns a probability for each level directly, without generating text, so the expected relevance can be read off the returned distribution. TypeSafe states that \jev{} is trained to return calibrated decisions~\cite{typesafe2026jev}. If the promise holds, \jev{} supplies exactly the calibrated relevance estimates that \emph{Beyond Yes and No} left open.

So far the promise has been checked only indirectly. Public benchmarks compare \jev{}'s rankings with cross-encoders~\cite{carlaiau2026jevreranking,emretheus2026jevrag}. One of them also reports the calibration of its relevance probabilities, but against binary SciFact and XQuAD labels rather than graded human judgments~\cite{emretheus2026jevrag}. Outside ranking, an independent calibration audit on text classification finds that \jev's calibration depends on the task and the question form~\cite{meng2026audit}. The question is also not tied to one closed service, since open-weight decision models with the same interface have since appeared~\cite{cloudflare2026clef,perplexity2026decider}.

We therefore ask:

\begin{itemize}
\item \textbf{RQ1:} do the ranking findings of \emph{Beyond Yes and No} reproduce, and do they carry over to decision models (Section~\ref{sec:rq1})?
\item \textbf{RQ2:} are the relevance probabilities calibrated against human judgments, natively and after a two-parameter calibration, and does that calibration transfer to a new year, corpus and domain (Section~\ref{sec:rq2})?
\item \textbf{RQ3:} what decisions do calibrated probabilities support that a ranking score does not (Section~\ref{sec:rq3})?
\end{itemize}

\section{Experimental Setup}\label{sec:setup}

\subsection{Data and text}

We use the eight BEIR collections of the reproduced paper~\cite{thakur2021beir}: TREC-COVID, Touch\'e-2020, DBPedia-Entity, SciFact, Signal-1M, TREC-NEWS, Robust04 and NFCorpus. BM25 top-100 candidates come from \texttt{castorini/rank\_llm\_data}, the RankGPT lineage~\cite{sun2023rankgpt}; qrels come from Anserini/Pyserini~\cite{yang2017anserini,lin2021pyserini}. As a check on these candidates, BM25 alone matches the reproduced paper on seven collections (six exactly, Signal-1M within 0.0004); NFCorpus yields 0.3218 versus the paper's 0.3075, which we could not trace.

For calibration, we add TREC DL19/20/21~\cite{craswell2020dl19,craswell2021dl20,craswell2022dl21}, which use the same BM25 top-100 setup and judge both relevant and non-relevant passages on a four-point scale. DL19 is the only split used for fitting, DL20 is development, and DL21 is the held-out test split, with a new year and corpus (MS MARCO v2). Only explicitly judged pairs enter probability analyses, since an unjudged pair has no human label; these number 2,257 / 2,859 / 2,786 pairs across 43 / 54 / 53 queries, respectively.

All scorers read the same text, title plus body, truncated to the first 256 \flan{} tokens, which the cross-encoders tokenize with their own tokenizers (maximum 512 tokens). Because the paper does not report truncation, Section~\ref{sec:rq1} also reports a 512-token rerun.

\subsection{Scorers}\label{sec:scorers}

\textbf{\flan{}.} BF16 inference uses the prompt templates reported in the reproduced paper. Teacher-forced label log-likelihoods exclude EOS in the primary score. Generated uses greedy continuation of at most 8 tokens, parsed to a label.

\textbf{\jev{}.} \texttt{typesafe/jev-1.13-20260917} via OpenRouter receives a \texttt{score} question with the same instruction and ordered labels; ranking uses ER over the returned distribution. RG-S is omitted because the API judges levels from descriptions. Each question gets one request; every reported \jev{} number is one draw. To check that one draw suffices, we reissued the DL21 request three more times for 300 judged pairs; this changed event probability by 0.010 on average and Brier score by at most 0.004, smaller than most differences between models.

\textbf{Cross-encoders.} We use MiniLM-L6, bge-reranker-v2-m3 and monoT5-3B~\cite{nogueira2020monot5}, the paper's own comparator, whose score is its true/false log-softmax.

\textbf{Open-weight decision models.} Clef-flash 9B~\cite{cloudflare2026clef}\footnote{\url{https://huggingface.co/Cloudflare/clef-flash}} and pplx-decider-v1.1 27B~\cite{perplexity2026decider}\footnote{\url{https://huggingface.co/perplexity-ai/pplx-decider-v1.1-27b}} use their vendors' release code, BF16, one question per request, and the same views, questions and splits as \jev{}.

\subsection{Calibration and evaluation}

\textbf{Relevance event and native probabilities.} A judged TREC pair counts as relevant if its human grade is $\geq 2$, TREC's binary relevance level; grades 0 (irrelevant) and 1 (related but not answering) are the negatives, and unjudged candidates are excluded rather than treated as negatives. Each model gives a native estimate $p$ of the probability of this event: $p_2+p_3$ of its four-grade distribution for \flan{} and the decision models, and the sigmoid of the logit for the cross-encoders.

\textbf{Calibration maps.} A calibration map turns a native probability into a corrected one, using human labels from a fitting split. With $x=\mathrm{logit}(p)$ and $p$ clipped to $[10^{-6}, 1-10^{-6}]$, Platt scaling~\cite{platt1999probabilistic} fits two parameters, $p'=\sigma(ax+b)$ with $a>0$: the slope $a$ spreads or squeezes the probabilities and the intercept $b$ moves them all up or down. The map is fitted by maximum likelihood on the DL19 labels and applied unchanged to DL20 and DL21. It is monotone, so it never changes a ranking.

\textbf{Metrics.} The Brier score~\cite{brier1950verification} is the mean squared difference between the probability and the 0/1 label; log loss is the mean negative log-likelihood (probabilities clipped at $10^{-6}$) and punishes confident mistakes most; ECE~\cite{naeini2015ece} bins the predictions into ten bins and averages the gap between the mean prediction and the observed rate. AUC, the chance that a random relevant pair is scored above a random non-relevant one, measures ordering only and is unchanged by calibration. Ranking is measured by nDCG@10 over complete qrels, with ties kept in BM25 order. 
Paired differences are tested with a 10,000-draw query bootstrap; `significant' means the 95\% interval excludes zero.

\section{Ranking: Reproduction and Decision Models (RQ1)}\label{sec:rq1}

\subsection{Reproducing Beyond Yes and No}

Table~\ref{tab:flan} compares our \flan{} runs with the paper's Table~6. All three findings reproduce in direction, but the absolute scores fall short.

\begin{table}[t]
\centering
\footnotesize
\setlength{\tabcolsep}{2.2pt}
\caption{nDCG@10 of \flan{} on the eight collections. ER columns give our score with the paper's Table~6 value in brackets~\cite{zhuang2024beyond}; Generated uses BM25-stable ties. Column heads abbreviate RG-$k$L and RG-S$(0,k)$.}
\label{tab:flan}
\resizebox{\textwidth}{!}{%
\begin{tabular}{@{}lrrrrrrrr@{}}
\toprule
 & \multicolumn{5}{c}{ER, ours [paper]} & \multicolumn{3}{c}{Generated} \\
\cmidrule(lr){2-6}\cmidrule(l){7-9}
Collection & 2L & 3L & 4L & S$(0,2)$ & S$(0,4)$ & 3L & 4L & S$(0,4)$ \\
\midrule
COVID & 0.749\,{\scriptsize[.777]} & 0.771\,{\scriptsize[.800]} & 0.774\,{\scriptsize[.803]} & 0.766\,{\scriptsize[.792]} & 0.771\,{\scriptsize[.797]} & 0.649 & 0.640 & 0.769 \\
Touch\'e & 0.260\,{\scriptsize[.274]} & 0.271\,{\scriptsize[.256]} & 0.261\,{\scriptsize[.248]} & 0.281\,{\scriptsize[.255]} & 0.274\,{\scriptsize[.264]} & 0.364 & 0.351 & 0.393 \\
DBPedia & 0.400\,{\scriptsize[.405]} & 0.430\,{\scriptsize[.430]} & 0.429\,{\scriptsize[.434]} & 0.422\,{\scriptsize[.425]} & 0.437\,{\scriptsize[.433]} & 0.371 & 0.372 & 0.401 \\
SciFact & 0.482\,{\scriptsize[.563]} & 0.642\,{\scriptsize[.701]} & 0.676\,{\scriptsize[.719]} & 0.710\,{\scriptsize[.734]} & 0.710\,{\scriptsize[.739]} & 0.687 & 0.682 & 0.712 \\
Signal & 0.285\,{\scriptsize[.282]} & 0.295\,{\scriptsize[.293]} & 0.309\,{\scriptsize[.305]} & 0.303\,{\scriptsize[.300]} & 0.316\,{\scriptsize[.313]} & 0.329 & 0.330 & 0.339 \\
News & 0.447\,{\scriptsize[.457]} & 0.491\,{\scriptsize[.470]} & 0.492\,{\scriptsize[.471]} & 0.477\,{\scriptsize[.470]} & 0.478\,{\scriptsize[.456]} & 0.395 & 0.395 & 0.436 \\
Robust04 & 0.516\,{\scriptsize[.542]} & 0.541\,{\scriptsize[.558]} & 0.544\,{\scriptsize[.558]} & 0.549\,{\scriptsize[.550]} & 0.549\,{\scriptsize[.545]} & 0.406 & 0.406 & 0.467 \\
NFCorpus & 0.346\,{\scriptsize[.376]} & 0.353\,{\scriptsize[.376]} & 0.354\,{\scriptsize[.378]} & 0.345\,{\scriptsize[.370]} & 0.346\,{\scriptsize[.371]} & 0.321 & 0.321 & 0.331 \\
\midrule
Average & 0.436\,{\scriptsize[.459]} & 0.474\,{\scriptsize[.485]} & 0.480\,{\scriptsize[.489]} & 0.482\,{\scriptsize[.487]} & 0.485\,{\scriptsize[.490]} & 0.440 & 0.437 & 0.481 \\
\bottomrule
\end{tabular}%
}
\end{table}

\textbf{Claim 1, graded labels beat binary labels: reproduced.} Every graded setting beats RG-2L on average, by 0.038--0.049 against 0.026--0.030 in the paper (RG-3L: 0.038); our gains are larger because our RG-2L is the setting furthest below its reported score. The two textual settings win on all eight collections and the two numeric scales on seven. As in the paper, the gain comes mostly from SciFact (RG-3L $-$ RG-2L: 0.160 against the paper's 0.138) and is close to zero on NFCorpus (0.006 against 0.0001); on Touch\'e the paper's small loss ($-$0.018) becomes a small gain (0.011).

\textbf{Claim 2, likelihood scores beat generated labels: reproduced, with a caveat.} \flan{} generates the lowest label for almost every pair (4,793 of 5,000 on TREC-COVID), so a Generated ranking is mostly ties, and its score depends on how ties are broken, which the paper does not report. We therefore report two tie-breaking rules. With ties kept in BM25 order, ER leads Generated by 0.03--0.04 for the two textual label sets, but almost not at all for RG-S$(0,4)$ (0.485 against 0.481), whose generated ratings are spread over several levels. With ties in uniformly random order (expected nDCG@10), Generated falls to 0.20--0.36 against 0.47--0.48 for ER, a larger gap than the paper's Table~3 reports for FLAN PaLM2 S (0.40--0.44 against 0.50).

\textbf{Claim 3, gains flatten beyond three or four levels: reproduced up to four levels.} RG-4L adds 0.006 over RG-3L (paper 0.004) and RG-S$(0,4)$ adds 0.003 over RG-S$(0,2)$ (paper 0.003); the four graded settings lie within 0.011 of each other, as in the paper's Table~6. Scoring by the likelihood of the highest label (PR) instead of ER stays within 0.01, also as in the paper.

\textbf{Absolute scores: not reproduced.} Our averages are 0.005--0.024 below the paper's. The gap differs by collection: within 0.005 on DBPedia and Signal-1M, where our 256-token view rarely truncates; 0.024--0.081 on SciFact, 0.025--0.029 on TREC-COVID and 0.023--0.029 on NFCorpus, where it truncates most documents, and TREC-NEWS even scores above the paper for every graded setting. Truncation is therefore a natural suspect, but rerunning these three collections with a 512-token view does not close the gap: SciFact RG-2L moves from 0.482 to 0.473 against the paper's 0.563, TREC-COVID gains 0.007--0.016 with intervals through zero, and NFCorpus gains 0.001--0.006. Since truncation does not explain the gap, the remaining candidates are details the paper does not report: preprocessing, prompt formatting, checkpoint and precision.

\subsection{Decision Models and Cross-Encoders}
We next rank the same candidates and text with the three decision models and the three cross-encoders (Table~\ref{tab:dm}).

\begin{table}[t]
\centering
\footnotesize
\setlength{\tabcolsep}{2.2pt}
\caption{nDCG@10 of cross-encoders and decision models on the shared text view. Decision models rank by ER; ties keep the BM25 order. pplx-decider is v1.1 (27B). The best value in each column is in bold.}
\label{tab:dm}
\resizebox{\textwidth}{!}{%
\begin{tabular}{@{}lrrrrrrrrr@{}}
\toprule
Method & COVID & Touch\'e & DBP. & SciF. & Signal & News & Rob04 & NFC. & Avg \\
\midrule
MiniLM-L6 & 0.753 & 0.285 & 0.443 & 0.681 & 0.340 & 0.450 & 0.453 & 0.351 & 0.469 \\
bge-reranker-v2-m3 & 0.802 & 0.347 & 0.445 & 0.719 & 0.290 & 0.454 & 0.489 & 0.339 & 0.486 \\
monoT5-3B & 0.793 & 0.316 & 0.448 & 0.757 & 0.320 & 0.495 & 0.544 & 0.370 & 0.505 \\
\addlinespace[3pt]
\jev{} RG-2L & 0.852 & \textbf{0.377} & 0.429 & 0.783 & 0.309 & 0.477 & 0.617 & 0.376 & 0.528 \\
\jev{} RG-3L & 0.854 & 0.313 & 0.430 & 0.786 & 0.307 & 0.476 & 0.614 & 0.380 & 0.520 \\
\jev{} RG-4L & 0.847 & 0.307 & 0.435 & 0.794 & 0.306 & 0.477 & 0.612 & 0.381 & 0.520 \\
\addlinespace[3pt]
Clef-flash RG-2L & 0.816 & 0.266 & 0.414 & 0.765 & 0.288 & 0.437 & 0.606 & 0.372 & 0.495 \\
Clef-flash RG-3L & 0.827 & 0.265 & 0.422 & 0.770 & 0.306 & 0.442 & 0.607 & 0.374 & 0.502 \\
Clef-flash RG-4L & 0.826 & 0.257 & 0.435 & 0.771 & 0.310 & 0.442 & 0.611 & 0.375 & 0.503 \\
\addlinespace[3pt]
pplx-decider RG-2L & \textbf{0.876} & 0.347 & 0.483 & 0.791 & 0.318 & 0.516 & \textbf{0.625} & 0.382 & 0.542 \\
pplx-decider RG-3L & 0.874 & 0.345 & 0.490 & \textbf{0.801} & 0.334 & \textbf{0.532} & 0.623 & \textbf{0.385} & \textbf{0.548} \\
pplx-decider RG-4L & 0.874 & 0.335 & \textbf{0.492} & 0.800 & \textbf{0.341} & 0.529 & 0.620 & 0.384 & 0.547 \\
\bottomrule
\end{tabular}%
}
\end{table}

\textbf{\jev{} ranks ahead of \flan{} and the cross-encoders on average.} \jev{} RG-3L beats MiniLM-L6 by 0.050, bge-reranker-v2-m3 by 0.034 and \flan{} RG-3L by 0.046. Its lead over monoT5-3B~\cite{nogueira2020monot5}, the strongest cross-encoder here, is small (0.014) and holds on four of the eight collections: it is large on Robust04 and TREC-COVID and absent on DBPedia, Touch\'e, Signal-1M and TREC-NEWS.

\textbf{The open-weight decision models bracket \jev{}.} The 9B Clef-flash~\cite{cloudflare2026clef} RG-3L is below \jev{} on every collection ($-0.018$) and level with monoT5-3B, whereas the 27B pplx-decider-v1.1~\cite{perplexity2026decider} RG-3L is above \jev{} on all eight collections (0.028) and above monoT5-3B by 0.042.

\textbf{Graded labels barely help decision models.} Moving from RG-2L to RG-3L adds 0.038 for \flan{} but only about 0.006 for Clef-flash and pplx-decider, and \jev{} loses 0.008. \jev{}'s small loss depends on tie-breaking: its two-decimal probabilities leave many documents tied, and breaking those ties at random instead of by BM25 order turns the loss into a gain of 0.011. For decision models the graded-label gain is therefore small and model-dependent.

\textbf{Answer to RQ1.} The three ranking findings reproduce in direction with \flan{}, although the margin of likelihood scores over generated labels depends on tie-breaking and our absolute scores stay below the paper's. \jev{} and the open 27B pplx-decider rank ahead of \flan{} and all three cross-encoders, but finer labels add little for them. Sections~\ref{sec:rq2} and~\ref{sec:rq3} turn from their rankings to their probabilities.

\section{Are the Probabilities Calibrated, and Does Calibration Transfer? (RQ2)}\label{sec:rq2}

Section~\ref{sec:rq1} used the prompts of \emph{Beyond Yes and No} to stay comparable with the original paper. Calibration, however, is judged against human assessors, so \flan{} and the decision models answer the question the TREC assessors answered: the TREC Deep Learning four-point scale with its official grade definitions (\emph{Irrelevant}: the passage has nothing to do with the query; \emph{Related}: it seems related but does not answer it; \emph{Highly relevant}: it has some answer, possibly unclear or hidden among extraneous information; \emph{Perfectly relevant}: it is dedicated to the query and contains the exact answer), framed as a NIST assessor's task, with the reminder that ``Related'' is not relevant. The end of this section tests how much this choice of wording matters. With this question, we compare native and DL19-calibrated probabilities (Section~\ref{sec:setup}) with the human labels of all judged pairs: DL20 checks the map on the same corpus, DL21 on a new year and corpus, and four BEIR collections across domains.

\begin{figure}[t]
  \centering
  \includegraphics[width=\textwidth]{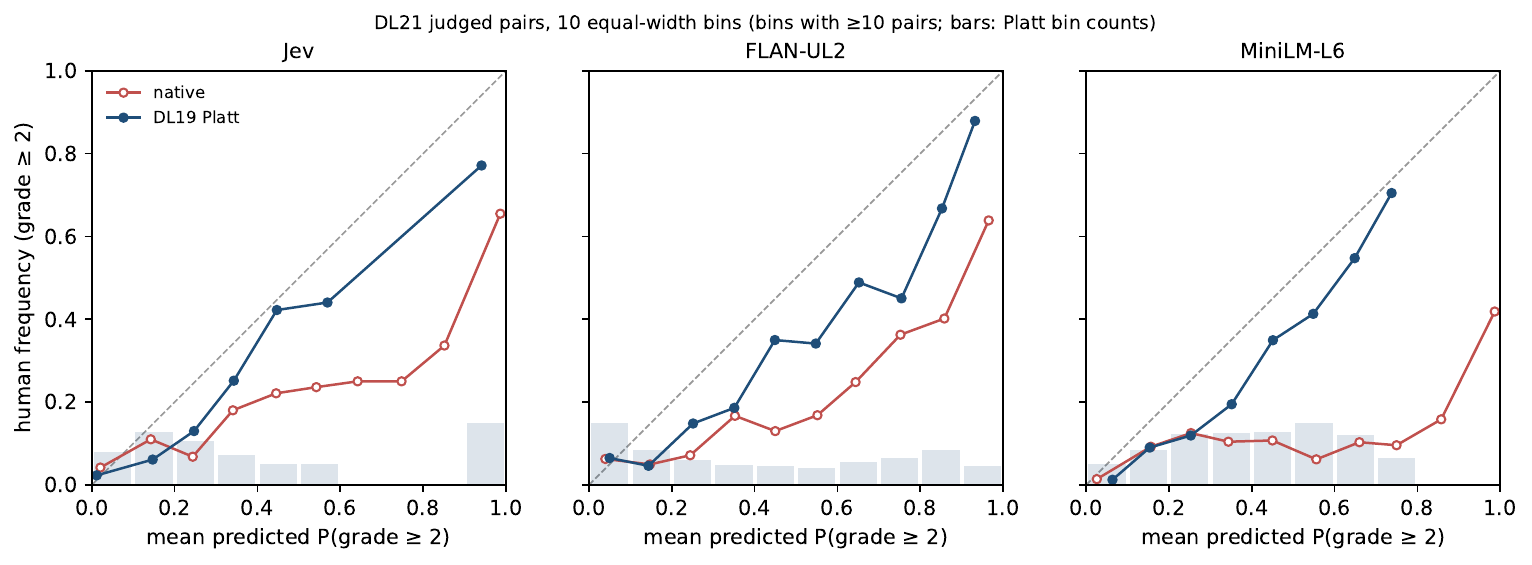}
  \caption{Reliability diagrams on DL21 judged pairs (event: human grade $\geq 2$), native probabilities versus the DL19-fitted Platt calibration.}
  \label{fig:reliability}
\end{figure}

\begin{table}[t]
\centering
\footnotesize
\setlength{\tabcolsep}{4pt}
\caption{Calibration against human judgments on explicitly judged pairs. Native probabilities and the Platt map fitted on DL19 only, applied unchanged to DL20, DL21 and four BEIR collections (TREC-COVID, DBPedia-Entity, TREC-NEWS, Robust04; unweighted mean). Event: grade $\geq 2$ on TREC, $\geq 1$ on BEIR. \flan{} and the decision models are asked the TREC assessor question; the open decision models were not run on BEIR. Lower is better except for AUC, which the monotone Platt map leaves unchanged; the best value per block and column is in bold.}
\label{tab:calib}
\begin{tabular}{@{}lrrrrrrr@{}}
\toprule
 & \multicolumn{2}{c}{Brier} & \multicolumn{2}{c}{Log loss} & \multicolumn{2}{c}{ECE$_{10}$} & \\
\cmidrule(lr){2-3}\cmidrule(lr){4-5}\cmidrule(lr){6-7}
Model & native & Platt & native & Platt & native & Platt & AUC \\
\midrule
\multicolumn{8}{@{}l}{\textit{DL20 (development, same corpus as DL19)}} \\
\flan{} & 0.186 & 0.133 & 0.619 & 0.407 & 0.204 & 0.112 & 0.872 \\
MiniLM-L6 & 0.403 & 0.145 & 1.944 & 0.446 & 0.451 & 0.135 & 0.837 \\
bge-reranker-v2-m3 & 0.199 & 0.128 & 0.691 & 0.394 & 0.212 & 0.105 & 0.869 \\
monoT5-3B & 0.234 & 0.143 & 0.822 & 0.430 & 0.246 & 0.132 & 0.857 \\
\jev{} & 0.185 & 0.123 & 1.096 & 0.410 & 0.198 & 0.118 & 0.886 \\
Clef-flash & \textbf{0.152} & 0.119 & \textbf{0.474} & 0.375 & 0.203 & 0.095 & 0.891 \\
pplx-decider & 0.159 & \textbf{0.114} & 0.503 & \textbf{0.358} & \textbf{0.188} & \textbf{0.090} & \textbf{0.903} \\
\midrule
\multicolumn{8}{@{}l}{\textit{DL21 (held-out test, new year and corpus)}} \\
\flan{} & 0.237 & 0.171 & 0.783 & 0.514 & 0.248 & 0.124 & 0.831 \\
MiniLM-L6 & 0.476 & 0.188 & 2.373 & 0.554 & 0.503 & 0.108 & 0.766 \\
bge-reranker-v2-m3 & 0.280 & 0.177 & 0.984 & 0.523 & 0.291 & 0.121 & 0.811 \\
monoT5-3B & 0.327 & 0.193 & 1.160 & 0.557 & 0.359 & 0.177 & 0.823 \\
\jev{} & 0.213 & 0.147 & 1.315 & 0.478 & \textbf{0.211} & 0.100 & 0.861 \\
Clef-flash & 0.202 & 0.172 & \textbf{0.596} & 0.519 & 0.212 & 0.115 & 0.836 \\
pplx-decider & \textbf{0.192} & \textbf{0.144} & 0.611 & \textbf{0.447} & 0.212 & \textbf{0.094} & \textbf{0.878} \\
\midrule
\multicolumn{8}{@{}l}{\textit{BEIR, mean of four collections (new domains)}} \\
\flan{} & \textbf{0.177} & 0.199 & \textbf{0.578} & 0.611 & \textbf{0.124} & 0.179 & 0.805 \\
MiniLM-L6 & 0.241 & 0.217 & 1.251 & 0.685 & 0.227 & 0.181 & 0.746 \\
bge-reranker-v2-m3 & 0.215 & 0.216 & 0.938 & 0.703 & 0.186 & 0.191 & 0.771 \\
monoT5-3B & 0.204 & \textbf{0.175} & 0.902 & \textbf{0.552} & 0.177 & \textbf{0.095} & 0.781 \\
\jev{} & 0.197 & 0.214 & 1.044 & 0.654 & 0.175 & 0.206 & \textbf{0.807} \\
\bottomrule
\end{tabular}
\end{table}

\textbf{Native probabilities are not calibrated; a Platt map fitted on DL19 fixes much of it.} Figure~\ref{fig:reliability} groups the DL21 pairs into ten bins by predicted probability and plots how often the pairs in each bin are actually relevant; a calibrated model lies on the diagonal. The native curves of all three models lie well below it, which means that every model is too optimistic. In particular, of the pairs to which \jev{} gives more than 0.9, only 66\% are relevant (\flan{} 64\%, MiniLM 42\%). \jev{} is also the only model that returns exactly 0 or 1, for 1,006 of the 2,786 pairs, 158 of them on the wrong side of the human label, so its native log loss (1.315) is worse than \flan{}'s (0.783) although its Brier score and AUC are slightly better (neither difference significant). By contrast, the open decision models are much less extreme (native log loss 0.596 and 0.611). After the Platt map fitted on DL19, the curves move towards the diagonal, and in Table~\ref{tab:calib} every model's Brier score, log loss and ECE drop on both DL20 and DL21; for \jev{} on DL21, the Brier score falls from 0.213 to 0.147, log loss from 1.315 to 0.478 and ECE from 0.211 to 0.100. The correction is not complete: on DL21 the calibrated curves still lie somewhat below the diagonal at high probabilities.

\textbf{After calibration, \jev{} has a lower DL21 Brier score than \flan{} and the cross-encoders.} Against \flan{}, its paired $\Delta$Brier is $-0.023$, while its $\Delta$log loss of $-0.036$ is not significant; against MiniLM, $-0.041$ and $-0.076$. The stronger cross-encoders also trail: relative to \jev{}, bge-reranker-v2-m3 has $\Delta$Brier $+0.030$ and monoT5-3B $+0.046$. On DL20, however, bge and \jev{} are indistinguishable ($\Delta$Brier $+0.005$), so the DL21 margin over the cross-encoders is one year's result. These two cross-encoders were added after the DL21 results of the first three models were known, but their calibrators were fitted on DL19 by the same code and their DL21 predictions had not been seen before evaluation.

\textbf{The open 27B decision model matches \jev{}; the 9B one does not hold up on the new corpus.} pplx-decider has the lowest calibrated DL21 Brier and log loss: 0.144 against \jev{}'s 0.147 and 0.447 against 0.478, neither difference significant. The 9B Clef-flash, in contrast, matches \jev{} on DL20 (Brier 0.119 versus 0.123) but falls to \flan{}'s level on DL21 (0.172 versus 0.171), $+0.025$ behind \jev{}.

\textbf{Across domains, the DL19 map still lowers log loss, but the Brier score depends on the collection.} The last block of Table~\ref{tab:calib} applies the DL19 map unchanged to four BEIR collections that have explicit non-relevant judgments (relevance: grade $\geq 1$) and averages over them. The map lowers the log loss of \jev{} and the three cross-encoders on every collection; \jev{}'s mean falls from 1.044 to 0.654. The Brier score, however, is mixed. \jev{}'s improves on TREC-NEWS and, barely, on Robust04 but worsens on TREC-COVID and DBPedia-Entity, so its mean rises from 0.197 to 0.214, and its ECE follows the same pattern. TREC-COVID is the hardest case, where the map raises the Brier score of every model except monoT5-3B, the only model whose Brier score improves on all four collections. For \flan{}, whose native probabilities are already moderate with this question, the map is unnecessary and raises the Brier score on all four collections and the log loss on all but Robust04. \jev{} and \flan{} also discriminate almost equally well across domains. \jev{} has the highest mean AUC (0.807; \flan{} 0.805), leading on TREC-COVID and Robust04, while \flan{} leads on DBPedia-Entity and TREC-NEWS.

\begin{table}[t]
\centering
\footnotesize
\setlength{\tabcolsep}{4pt}
\caption{\jev{} on the same DL21 pairs with the TREC assessor question and with the original paper's two-level question. Platt maps are fitted per question on DL19; $\rho$ is the Spearman correlation with the TREC assessor question's probabilities.}
\label{tab:prompts}
\begin{tabular}{@{}lrrrrrr@{}}
\toprule
 & \multicolumn{2}{c}{Brier} & \multicolumn{2}{c}{Log loss} & & \\
\cmidrule(lr){2-3}\cmidrule(lr){4-5}
Question & native & Platt & native & Platt & AUC & $\rho$ \\
\midrule
TREC assessor question, four levels& 0.213 & 0.147 & 1.315 & 0.478 & 0.861 & -- \\
Original two-level question (RG-2L) & 0.322 & 0.156 & 1.863 & 0.507 & 0.855 & 0.921 \\
\bottomrule
\end{tabular}
\end{table}

\textbf{The remaining advantage is mainly discrimination.} Returning to DL21, after Platt scaling the models' ECE values are close (0.094--0.124, except 0.177 for monoT5-3B), so apart from monoT5-3B their probability scales are similarly accurate. What still differs is AUC, which calibration cannot change: 0.766 for MiniLM, 0.861 for \jev{} and 0.878 for pplx-decider. \jev{} leads MiniLM by $+0.095$ and \flan{} by $+0.030$ (not significant). Hence the two models with the highest AUC, pplx-decider and \jev{}, also have the best calibrated Brier scores. 
In ranking, too, on DL21 \jev{}'s nDCG@10 of 0.705 is ahead of MiniLM (0.655, $+0.050$), bge (0.662, $+0.043$) and monoT5 (0.668, $+0.037$), and pplx-decider reaches 0.726 ($+0.021$ over \jev{}).

\textbf{The wording changes \jev's native probabilities, hardly its calibrated ones.} Asked the original paper's two-level question instead of the assessor question, \jev's native DL21 Brier score rises from 0.213 to 0.322 and its log loss from 1.315 to 1.863; after a Platt map fitted on DL19 for each question, the two are close (0.147 and 0.156), with similar AUC and highly correlated probabilities (Table~\ref{tab:prompts}). The same holds for six further wordings (level names only, no assessor role, a paraphrase, a choice question and two direct binary questions), which give native Brier scores from 0.196 to 0.308 but all lie within $+0.006$ of the assessor question's Brier score after calibration. \jev{} follows the question it is given, but the difference lies mostly in the probability scale, which calibration removes.

\textbf{Answer to RQ2.} On TREC, no model's native output can be read as a calibrated probability of human relevance, and only \jev{} returns exact 0s and 1s. However, a two-parameter calibration fitted on DL19 corrects much of this and transfers to DL20 and DL21. After calibration, \jev{} and the open pplx-decider give the best probabilities on the held-out year, about equally, and their advantage comes mainly from separating relevant from non-relevant pairs better rather than from a more accurate probability scale. \jev{}'s calibrated probabilities hardly depend on the question's wording. Across domains, the frozen DL19 map still lowers the log loss of \jev{} and the cross-encoders on every BEIR collection, but not of \flan{}; the Brier score improves on some collections and worsens on others. Section~\ref{sec:rq3} tests what decisions these probabilities support.

\begin{table}[t]
\centering
\footnotesize
\setlength{\tabcolsep}{4pt}
\caption{Mean F1 of truncated lists on DL21 (judged documents, relevance grade $\geq 2$). Fixed cut-offs, a threshold tuned on DL19, the expected-F1 rule with native and with Platt probabilities, and the oracle cut-off.}
\label{tab:truncation}
\begin{tabular}{@{}lrrrrrr@{}}
\toprule
 & \multicolumn{2}{c}{Fixed $k$} & DL19 & \multicolumn{2}{c}{Expected F1} & \\
\cmidrule(lr){2-3}\cmidrule(lr){5-6}
Model & 10 & 20 & threshold & native & Platt & Oracle \\
\midrule
\jev{} & 0.506 & 0.544 & 0.630 & 0.618 & 0.638 & 0.778 \\
\flan{} & 0.439 & 0.503 & 0.561 & 0.548 & 0.567 & 0.725 \\
MiniLM-L6 & 0.440 & 0.478 & 0.524 & 0.476 & 0.517 & 0.733 \\
bge-reranker-v2-m3 & 0.443 & 0.511 & 0.557 & 0.549 & 0.566 & 0.743 \\
monoT5-3B & 0.451 & 0.506 & 0.566 & 0.525 & 0.530 & 0.741 \\
Clef-flash & 0.464 & 0.519 & 0.576 & 0.556 & 0.593 & 0.744 \\
pplx-decider & 0.503 & 0.552 & 0.643 & 0.614 & 0.658 & 0.790 \\
\bottomrule
\end{tabular}
\end{table}

\section{What Do Calibrated Probabilities Buy? (RQ3)}\label{sec:rq3}

A calibrated probability supports decisions that a ranking score cannot, such as where to cut a ranked list. As in Section~\ref{sec:rq2}, all parameters are fitted on DL19 and tested on DL21.

\textbf{Expected-F1 truncation.} Ranked-list truncation chooses, for each query, how many top documents to return~\cite{lien2019bicut,bahri2020choppy}. The choice matters whenever each returned document has a cost, such as a reader's attention, a slot for an expensive LLM re-ranker~\cite{meng2024rlt}, or context tokens in retrieval-augmented generation~\cite{lewis2020rag}, where irrelevant passages can also distract the model~\cite{shi2023distracted}. Let $p_i$ be the calibrated probability that the document at rank $i$ is relevant. Then $\sum_{i\leq k}p_i$ is the expected number of relevant documents in the top $k$ and $\sum_i p_i$ the expected number of relevant candidates; substituting both into $F_1=2\,\mathit{TP}/(k+R)$, where $\mathit{TP}$ counts the relevant documents returned and $R$ all relevant ones, gives the depth
\[
k^\star=\arg\max_k \frac{2\sum_{i\leq k}p_i}{k+\sum_i p_i},
\]
a plug-in form of predicting with maximum expected F1 from probabilities~\cite{ye2012optimizing}. Adding the next document raises this value only if its probability exceeds half of it, which matches the result that the F1-optimal threshold for calibrated probabilities is half the optimal F1~\cite{lipton2014thresholding}. The depth therefore adapts to each query without a tuned cut-off, provided the probabilities are calibrated.

Table~\ref{tab:truncation} compares this rule with three baselines on condensed lists (unjudged documents removed before cutting): fixed depths; a probability threshold $\tau$, which keeps documents down the ranking until the first one with $p_i<\tau$, with $\tau$ chosen on DL19 to maximise mean F1 over a grid from 0.05 to 0.95 (0.30 for calibrated \jev{}); and an oracle that picks each query's best depth from the human labels. Calibrated \jev{} reaches F1 0.638, level with the tuned threshold (0.630) and well above a fixed cut at $k=10$ (0.506), though still below the oracle (0.778). The rule also beats the tuned threshold for \flan{}, bge and both open decision models, but not for MiniLM and monoT5. With calibrated probabilities, \jev{} and pplx-decider also cut better than all three cross-encoders (0.638 and 0.658 against at most 0.566). Calibration raises F1 for every model, since the overly optimistic native probabilities make the rule cut too deep; \jev{} improves from 0.618, while its mean depth falls from 28.6 to 26.1.

\textbf{Answer to RQ3.} Calibrated probabilities support list truncation without a tuned cut-off: the expected-F1 rule matches or beats a threshold tuned on DL19 for five of the seven models, and calibration improves it for every model. With this rule, \jev{} and pplx-decider cut better than every cross-encoder.

\section{Conclusion}

We reproduced \emph{Beyond Yes and No} with \flan{} on eight BEIR collections: its three ranking claims hold in direction, although our absolute scores fall short of the paper's. \jev{} and the open pplx-decider rank ahead of \flan{} and three cross-encoders on average, but finer relevance labels add little for decision models. Ranking quality, however, says little about probability quality. Against TREC assessors, no model's native probabilities are calibrated, including \jev{}'s despite its vendor's promise. Still, a two-parameter Platt map fitted on DL19 transfers to DL20 and DL21, after which \jev{} and pplx-decider give the best probabilities on DL21, mainly through better discrimination; across domains the map still lowers the log loss of \jev{} and the cross-encoders, but not always the Brier score. Calibrated probabilities in turn support list truncation without a tuned cut-off: an expected-F1 rule matches or beats a cut-off threshold tuned on DL19 for five of the seven models, and with it \jev{} and pplx-decider cut better than every cross-encoder.

\bibliographystyle{splncs04}
\bibliography{refs}

\begin{thebibliography}{10}
\providecommand{\url}[1]{\texttt{#1}}
\providecommand{\urlprefix}{URL }
\providecommand{\doi}[1]{https://doi.org/#1}

\bibitem{bahri2020choppy}
Bahri, D., Tay, Y., Zheng, C., Metzler, D., Tomkins, A.: Choppy: Cut
  transformer for ranked list truncation. In: Proceedings of the 43rd
  International ACM SIGIR Conference on Research and Development in Information
  Retrieval. pp. 1513--1516 (2020)

\bibitem{brier1950verification}
Brier, G.W.: Verification of forecasts expressed in terms of probability.
  Monthly Weather Review  \textbf{78}(1) (1950)

\bibitem{carlaiau2026jevreranking}
{carlaiau}: {JEV} reranking comparisons. GitHub repository,
  \url{https://github.com/carlaiau/jev-reranking} (2026)

\bibitem{cloudflare2026clef}
{Cloudflare}: {Clef-Flash}. Hugging Face model card,
  \url{https://huggingface.co/Cloudflare/clef-flash} (2026), revision 17f0b0ad,
  Apache-2.0

\bibitem{elastic2026jev}
Coates, D.: Reranking search results with a {System One} model: Testing {Jev}.
  \url{https://www.elastic.co/search-labs/blog/ecommerce-search-reranking-llm-alternative-jev}
  (2026), blog post, 30 September 2026

\bibitem{craswell2021dl20}
Craswell, N., Mitra, B., Yilmaz, E., Campos, D.: Overview of the {TREC} 2020
  deep learning track. In: Proceedings of TREC 2020 (2021)

\bibitem{craswell2022dl21}
Craswell, N., Mitra, B., Yilmaz, E., Campos, D., Lin, J.: Overview of the
  {TREC} 2021 deep learning track. In: Proceedings of the Thirtieth Text
  REtrieval Conference (TREC 2021). NIST Special Publication 500-335, National
  Institute of Standards and Technology (2021)

\bibitem{craswell2020dl19}
Craswell, N., Mitra, B., Yilmaz, E., Campos, D., Voorhees, E.M.: Overview of
  the {TREC} 2019 deep learning track. In: Proceedings of TREC 2019 (2020)

\bibitem{dinzinger2026system1}
Dinzinger, M., Ghosh~Dastidar, K., Granitzer, M.: {System One} models are not
  better rerankers. arXiv preprint arXiv:2610.32318  (2026)

\bibitem{dorosario2026jevrag}
Do~Rosario, A., Atef, K.: {Jev}-as-a-judge for {RAG} claim verification:
  Confidence-gated cascading between a non-generative verifier and a frontier
  {LLM}. arXiv preprint arXiv:2610.20634  (2026)

\bibitem{emretheus2026jevrag}
{emretheus}: {Jev} {RAG} benchmark ({English}): {Jev} 1.13 vs {NVIDIA}
  cross-encoder vs no reranking. GitHub repository,
  \url{https://github.com/emretheus/jev-rag-benchmark} (2026)

\bibitem{faggioli2023perspectives}
Faggioli, G., Dietz, L., Clarke, C.L.A., Demartini, G., Hagen, M., Hauff, C.,
  Kando, N., Kanoulas, E., Potthast, M., Stein, B., Wachsmuth, H.: Perspectives
  on large language models for relevance judgment. In: Proceedings of ICTIR
  (2023)

\bibitem{fernandes2026jevptbr}
Fernandes, L.C., Pereira, J.: {Jev}-as-a-judge for {Brazilian Portuguese}
  information retrieval: Relevance assessment and reranking. arXiv preprint
  arXiv:2610.23346  (2026)

\bibitem{guo2017calibration}
Guo, C., Pleiss, G., Sun, Y., Weinberger, K.Q.: On calibration of modern neural
  networks. In: Proceedings of ICML (2017)

\bibitem{lewis2020rag}
Lewis, P., Perez, E., Piktus, A., Petroni, F., Karpukhin, V., Goyal, N.,
  K{\"u}ttler, H., Lewis, M., Yih, W.t., Rockt{\"a}schel, T., Riedel, S.,
  Kiela, D.: Retrieval-augmented generation for knowledge-intensive {NLP}
  tasks. In: Advances in Neural Information Processing Systems (2020)

\bibitem{li2026system1rag}
Li, Q., Xie, Y.: Can {System-1}-style decision models replace dedicated
  rerankers in {RAG}? {A} systems evaluation. arXiv preprint arXiv:2610.34816
  (2026)

\bibitem{lien2019bicut}
Lien, Y.C., Cohen, D., Croft, W.B.: An assumption-free approach to the dynamic
  truncation of ranked lists. In: Proceedings of the 2019 ACM SIGIR
  International Conference on Theory of Information Retrieval. pp. 79--82
  (2019)

\bibitem{lin2021pyserini}
Lin, J., Ma, X., Lin, S.C., Yang, J.H., Pradeep, R., Nogueira, R.: Pyserini: A
  {Python} toolkit for reproducible information retrieval research with sparse
  and dense representations. In: Proceedings of the 44th International ACM
  SIGIR Conference on Research and Development in Information Retrieval. pp.
  2356--2362 (2021)

\bibitem{lipton2014thresholding}
Lipton, Z.C., Elkan, C., Naryanaswamy, B.: Optimal thresholding of classifiers
  to maximize {F1} measure. In: Machine Learning and Knowledge Discovery in
  Databases (ECML PKDD). pp. 225--239 (2014)

\bibitem{ma2024ranklama}
Ma, X., Wang, L., Yang, N., Wei, F., Lin, J.: Fine-tuning {LLaMA} for
  multi-stage text retrieval. In: Proceedings of the 47th International ACM
  SIGIR Conference on Research and Development in Information Retrieval (2024)

\bibitem{meng2024rlt}
Meng, C., Arabzadeh, N., Askari, A., Aliannejadi, M., de~Rijke, M.: Ranked list
  truncation for large language model-based re-ranking. In: Proceedings of the
  47th International ACM SIGIR Conference on Research and Development in
  Information Retrieval. pp. 141--151 (2024)

\bibitem{meng2026audit}
Meng, L.: How far can a commercial decision model's probabilities be trusted?
  {A} calibration audit of {Jev} against open and general-purpose classifiers.
  Preprint, eScholarship, University of California,
  \url{https://escholarship.org/content/qt4t4449jv/qt4t4449jv.pdf} (2026), 24
  September 2026

\bibitem{naeini2015ece}
Naeini, M.P., Cooper, G.F., Hauskrecht, M.: Obtaining well calibrated
  probabilities using {B}ayesian binning. In: Proceedings of the 29th AAAI
  Conference on Artificial Intelligence. pp. 2901--2907 (2015)

\bibitem{nogueira2020monot5}
Nogueira, R., Jiang, Z., Pradeep, R., Lin, J.: Document ranking with a
  pretrained sequence-to-sequence model. In: Findings of the Association for
  Computational Linguistics: EMNLP 2020. pp. 708--718 (2020)

\bibitem{perplexity2026decider}
{Perplexity AI}: pplx-decider-v1.1-27b. Hugging Face model card,
  \url{https://huggingface.co/perplexity-ai/pplx-decider-v1.1-27b} (2026),
  revision 14c9d901, Apache-2.0

\bibitem{platt1999probabilistic}
Platt, J.C.: Probabilistic outputs for support vector machines and comparisons
  to regularized likelihood methods. In: Advances in Large Margin Classifiers.
  MIT Press (1999)

\bibitem{pradeep2023rankvicuna}
Pradeep, R., Sharifymoghaddam, S., Lin, J.: {RankVicuna}: Zero-shot listwise
  document reranking with open-source large language models. arXiv:2309.15088
  (2023)

\bibitem{qin2024prp}
Qin, Z., Jagerman, R., Hui, K., Zhuang, H., Wu, J., Yan, L., Shen, J., Liu, T.,
  Liu, J., Metzler, D., Wang, X., Bendersky, M.: Large language models are
  effective text rankers with pairwise ranking prompting. In: Findings of the
  Association for Computational Linguistics: NAACL 2024. pp. 1504--1518 (2024)

\bibitem{shi2023distracted}
Shi, F., Chen, X., Misra, K., Scales, N., Dohan, D., Chi, E.H., Sch{\"a}rli,
  N., Zhou, D.: Large language models can be easily distracted by irrelevant
  context. In: Proceedings of the 40th International Conference on Machine
  Learning (2023)

\bibitem{sun2023rankgpt}
Sun, W., Yan, L., Ma, X., Wang, S., Ren, P., Chen, Z., Yin, D., Ren, Z.: Is
  {ChatGPT} good at search? {Investigating} large language models as re-ranking
  agents. In: Proceedings of EMNLP (2023)

\bibitem{tatar2026probrel}
Tatar, B., {\c{S}}ener, C.: Probabilistic relevance decisions for semantic
  retrieval. arXiv preprint arXiv:2611.08014  (2026)

\bibitem{terven2026scanindex}
Terven, J., C{\'o}rdova-Esparza, D.M., Romero-Gonz{\'a}lez, J.A.,
  Chavez-Urbiola, E., Vasquez, J.I., Ramirez-Bautista, J.A.,
  Chaparro-C{\'a}rdenas, S.L., Alfonso-Francia, G.: Scan or index? {Large}
  evidence coverage, answer quality, and cost in long-document {RAG}. arXiv
  preprint arXiv:2610.28957  (2026)

\bibitem{thakur2021beir}
Thakur, N., Reimers, N., R{\"u}ckl{\'e}, A., Srivastava, A., Gurevych, I.:
  {BEIR}: A heterogeneous benchmark for zero-shot evaluation of information
  retrieval models. In: NeurIPS Datasets and Benchmarks Track (2021)

\bibitem{tian2023calibration}
Tian, K., Mitchell, E., Zhou, A., Sharma, A., Rafailov, R., Yao, H., Finn, C.,
  Manning, C.D.: Just ask for calibration: Strategies for eliciting calibrated
  confidence scores from language models fine-tuned with human feedback. In:
  Proceedings of the 2023 Conference on Empirical Methods in Natural Language
  Processing (EMNLP) (2023)

\bibitem{typesafe2026jev}
{TypeSafe AI}: {TypeSafe} documentation. \url{https://docs.typesafe.ai/}
  (2026), accessed 2026-10-01; model typesafe/jev-1.13-20260917 via OpenRouter

\bibitem{wu2026decisionrank}
Wu, Y., Yu, D., Bai, C., Zhang, J., Liu, R., Chen, T., Mei, S., Cui, H., Shyu,
  F., Pandey, S., Simon, L., Liu, C., Liu, X.: {DecisionRank}: On-policy
  distillation for decision-model rerankers. arXiv preprint arXiv:2610.42128
  (2026)

\bibitem{wu2026sag2}
Wu, Y.: {SAG2}: Embedding-free multi-hop retrieval with {BM25} and {JEV}. arXiv
  preprint arXiv:2610.22588  (2026)

\bibitem{xiong2024uncertainty}
Xiong, M., Hu, Z., Lu, X., Li, Y., Fu, J., He, J., Hooi, B.: Can {LLM}s express
  their uncertainty? {An} empirical evaluation of confidence elicitation in
  {LLM}s. In: Proceedings of the Twelfth International Conference on Learning
  Representations (ICLR) (2024)

\bibitem{yang2017anserini}
Yang, P., Fang, H., Lin, J.: Anserini: Enabling the use of {Lucene} for
  information retrieval research. In: Proceedings of the 40th International ACM
  SIGIR Conference on Research and Development in Information Retrieval. pp.
  1253--1256 (2017)

\bibitem{ye2012optimizing}
Ye, N., Chai, K.M.A., Lee, W.S., Chieu, H.L.: Optimizing {F}-measures: A tale
  of two approaches. In: Proceedings of the 29th International Conference on
  Machine Learning. pp. 289--296 (2012)

\bibitem{zhuang2024beyond}
Zhuang, H., Qin, Z., Hui, K., Wu, J., Yan, L., Wang, X., Bendersky, M.: Beyond
  yes and no: Improving zero-shot {LLM} rankers via scoring fine-grained
  relevance labels. In: Proceedings of the 2024 Conference of the North
  American Chapter of the Association for Computational Linguistics: Human
  Language Technologies (Volume 2: Short Papers). pp. 358--370 (2024),
  arXiv:2310.14122

\bibitem{zhuang2023qlm}
Zhuang, S., Liu, B., Koopman, B., Zuccon, G.: Open-source large language models
  are strong zero-shot query likelihood models for document ranking. In:
  Findings of the Association for Computational Linguistics: EMNLP 2023. pp.
  8807--8817 (2023)

\bibitem{zhuang2026rankr1}
Zhuang, S., Ma, X., Yao, Z., Wang, S., Koopman, B., Lin, J., Zuccon, G.:
  {Rank-R1}: Enhancing reasoning in {LLM}-based document rerankers via
  reinforcement learning. In: Proceedings of the 49th International ACM SIGIR
  Conference on Research and Development in Information Retrieval (2026),
  arXiv:2503.06034

\bibitem{zhuang2024setwise}
Zhuang, S., Zhuang, H., Koopman, B., Zuccon, G.: A setwise approach for
  effective and highly efficient zero-shot ranking with large language models.
  In: Proceedings of the 47th International ACM SIGIR Conference on Research
  and Development in Information Retrieval (2024)

\bibitem{zuccon2025r2llms}
Zuccon, G., Zhuang, S., Ma, X.: {R$^2$LLMs}: Retrieval and ranking with {LLMs}.
  In: Proceedings of the 48th International ACM SIGIR Conference on Research
  and Development in Information Retrieval. pp. 4106--4109 (2025).
  \doi{10.1145/3726302.3731689}

\end{thebibliography}

\end{document}